\documentclass[twocolumn,showpacs,nofootinbib,amsmath,amssymb,aps,prd]{revtex4-1}
\usepackage{amsmath,amsfonts,bm}
\usepackage{graphicx}
\usepackage[colorlinks, linkcolor={red},citecolor={blue}]{hyperref}

\begin{document}
\title{Decoupling gravitational sources in\\general relativity: the extended case}
\author{J Ovalle}
\email[]{jovalle@usb.ve}
\affiliation{Institute of Physics and Research Centre of Theoretical Physics and Astrophysics, Faculty of Philosophy and Science, Silesian University in Opava, CZ-746 01 Opava, Czech Republic.\\Departamento de F\'{\i}sica, Universidad Sim\'on Bol\'{\i}var,
Apartado 89000, Caracas 1080A, Venezuela.}
\begin{abstract}
We show how to decoupling two spherically symmetric and static gravitational sources through the most general possible extension of the so-called Minimal Geometric Deformation-decoupling. As a test, we decouple the Einstein-Maxwell system and reproduce the Reissner-Nordstrom solution. We show the potential of this method to study i) the consequences of modified gravity on general relativity, ii) to investigate the conjectured dark matter, and iii) to study hairy black holes.
\end{abstract} 
\maketitle


\section{Introduction}
\setcounter{equation}{0}
In a recent paper~\cite{MGD-decoupling} we developed a simple, systematic and direct approach to decoupling gravitational sources in general relativity, the so-called Minimal Geometric Deformation-decoupling (MGD-decoupling, henceforth). We proved, contrary to the general belief, that it is possible to solve Einstein's field equations for a gravitational source, whose energy-momentum tensor $\tilde{T}_{\mu\nu}$ is expressed as
\begin{equation}
\label{emt}
\tilde{T}_{\mu\nu}
=
T^{\rm}_{\mu\nu}+\theta_{\mu\nu}
\ ,
\end{equation}
by solving Einstein's field equations for each component $\{T^{\rm}_{\mu\nu},\,\theta_{\mu\nu}\}$ separately. Then, by a straightforward superposition of the two  solutions, we obtain the complete solution corresponding to the source $\tilde{T}_{\mu\nu}$. Since Einstein's field equations are non-linear, the MGD-decoupling represents a novel and useful method in the search and analysis of solutions, especially when we face scenarios beyond trivial cases,
such as the interior of stellar systems with gravitational sources more
realistic than the ideal perfect fluid~\cite{lake2,visser2005}, or even when we consider alternative theories, which usually introduce new features difficult to deal with.
\par
The original version of the MGD approach was developed~\cite{jo1,jo2} in the context of the Randall-Sundrum brane-world~\cite{lisa1,lisa2} and extended to investigate new black hole solutions in Refs.~\cite{MGDextended1,MGDextended2}
(for some earlier works on the MGD, see for instance Refs.~\cite{jo6,jo8,jo9,jo10},
and Refs.~\cite{jo11,jo12,roldaoGL,rrplb,rr-glueball,rr-acustic,fr,MGDBH,emgd} for some recent applications). On the other hand, the MGD-decoupling, which is based on MGD, has a number of features that make it particularly attractive in the search for new solutions of Einstein's field equations~\cite{mgdaniso,Luciano,MGDDBH,sharif1,Inverse,sharif2,ernestopedro,tello,camilo,angel2,sharif3}. The two main characteristics of this approach~\cite{MGD-decoupling,mgdaniso} are: i) we can extend
simple solutions of the Einstein equations into more complex domains, that is to say, we can start from a simple gravitational source with energy-momentum tensor $T_{\mu\nu}$
and add to it a more complex gravitational source
\begin{equation}
\label{coupling0}
T_{\mu\nu}\mapsto \tilde T_{\mu\nu}=T_{\mu\nu}+T^{(1)}_{\mu\nu}
\ ,
\end{equation}
We can then repeat the process with more sources $T^{(i)}_{\mu\nu}$ to extend straightforward solutions of the Einstein equations, associated
with the simplest gravitational source $T_{\mu\nu}$, into the domain of more intricate forms of gravitational sources $\tilde T_{\mu\nu}$.
ii) We can also develop the converse of the above, that is, in order to find a solution to Einstein's equations with a complex energy-momentum tensor $\tilde{T}_{\mu\nu}$, we can split it into simpler components, namely
\begin{equation}
\label{split}
\tilde{T}_{\mu\nu}\rightarrow\{T_{\mu\nu},\,T^{(i)}_{\mu\nu}\}\ ,
\end{equation} 
and solve Einstein's equations for each one of these components. Hence, we will have as many solutions as are components in the original energy-momentum tensor $\tilde{T}_{\mu\nu}$.
Finally, by a simple combination of all these solutions, we will obtain the solution to the Einstein equations associated with the original energy-momentum tensor $\tilde{T}_{\mu\nu}$. We emphasize that the MGD-decoupling works as long as the sources do not exchange energy-momentum among them,
namely
\begin{equation}
\nabla_{\mu}T^{\mu\nu}
=
\nabla_{\mu} T^{(1)\mu\nu}
=
\ldots
=
\nabla_{\mu} T^{(n)\mu\nu}
=
0
\ ,
\label{nablas}
\end{equation}
which further clarifies that their interaction is purely gravitational. We want to emphasize that the MGD-decoupling is based in a specific transformation of the inverse of the radial metric component $g^{-1}_{rr}$, namely,
\begin{equation}
\hat{g}^{-1}_{rr}\rightarrow\,{g}^{-1}_{rr} = \hat{g}^{-1}_{rr}+\alpha\,f(r)\ ,
\end{equation}
where $f(r)$ represents the deformation undergone by the metric $\hat{g}_{\mu\nu}$ [see further~ Eq.~\eqref{gd2}]. This transformation is precisely what allows the decoupling of gravitational sources, and has been successfully used, among other things, to derive exact and physically acceptable solutions for spherically symmetric and non-uniform stellar distributions~\cite{jo8,jo9,mgdaniso,Luciano,Inverse,sharif1,tello,camilo,sharif3}; to study microscopic black holes~\cite{jo6};  to prove, contrary to previous claims, the consistency of a Schwarzschild exterior for a spherically symmetric self-gravitating system
made of regular matter in the brane-world~\cite{jo11};  to investigate the
gravitational lensing phenomena beyond general relativity~\cite{roldaoGL}, to determine the critical stability region for Bose-
Einstein condensates in gravitational systems~\cite{rrplb}, as well as to study the corrections to dark $SU(N)$
stars observable parameters due to variable tension fluid branes~\cite{rr-glueball}.
\par
While is true that the MGD-decoupling is a useful and powerful tool to investigate self-gravitating systems, it is fair to say that it has some limitations. Probably the main one is that the deformation undergone by the metric is minimal, that is, only the radial metric component $g_{rr}$ is deformed, leaving the temporal component $g_{tt}$ invariant. This could lead to certain drawbacks when we study, for instance,  the existence of stable black holes with a well-defined horizon~\cite{MGDDBH}. In this respect, the MGD approach, which represents the foundation of the MGD-decoupling, was successfully extended when both metric functions are deformed~\cite{MGDextended1,MGDextended2,emgd}. 
However, this extension works only in the vacuum and fails for regions where matter is present, since the Bianchi identities are no longer satisfied. This represents a disadvantage if we want to investigate, for instance, the interior of a self-gravitating distribution. Unfortunately, to develop a correct extension of this approach for regions where $T_{\mu\nu}\neq\,0$ represents a highly non-trivial task~\cite{MGDextended1}. The above indicates that the Bianchi identities play a preponderant role in the MGD-decoupling, and therefore we need to carry out a careful study on this identities to develop a successful extension of the MGD-decoupling (MGDe henceforth) for all regions of the space-time, regardless of whether there is matter or not. This is precisely the scenario under study in this paper.
\par
%
%
\par
The paper is organised as follows:
in Section~\ref{s2}, we successfully develop the decoupling of two spherically symmetric and static gravitational sources, $\{T_{\mu\nu},\,\theta_{\mu\nu}\}$ for all regions of the space-time, when both metric components $\{g_{tt},\,g_{rr}\}$ are deformed, showing in detail the critical role played by the Bianchi identities, as well as the potential application of the MGDe in extended theories; in Section~\ref{s3}, we test the consistence of the MGDe developed in Section~\ref{s2} by decoupling the Einstein-Maxwell system; finally, we summarise our conclusions in Section~\ref{con}.
\section{MGD decoupling for two sources}
\label{s2}
\par
Let us start from the standard Einstein field equations
\begin{equation}
\label{corr2}
R_{\mu\nu}-\frac{1}{2}\,R\, g_{\mu\nu}
=-
k^2\,\tilde{T}_{\mu\nu}
\ ,
\end{equation}
and assume the total energy-momentum tensor contains two contributions~\cite{Matt}, namely
\begin{equation}
\label{emt}
\tilde{T}_{\mu\nu}
=
T^{\rm}_{\mu\nu}+\theta_{\mu\nu}
\ ,
\end{equation}
These sources may contain new fields, like scalar, vector and tensor fields,
and will in general produce anisotropies in self-gravitating systems. Also, one of them could represent the effective energy-momentum tensor associated with a new gravitational sector [see further below Eq.~\eqref{ngt2}], non described by general relativity.
\par
Since the Einstein tensor satisfies the Bianchi identity, the total source in Eq.~(\ref{emt})
must satisfy the conservation equation
\begin{equation}
\nabla_\mu\,\tilde{T}^{\mu\nu}=0
\ .
\label{dT0}
\end{equation}
\par 
We next consider the spherically symmetric and static case. In Schwarzschild-like coordinates, a static spherically symmetric metric $g_{\mu\nu}$ reads 
\begin{equation}
ds^{2}
=
e^{\nu (r)}\,dt^{2}-e^{\lambda (r)}\,dr^{2}
-r^{2}\left( d\theta^{2}+\sin ^{2}\theta \,d\phi ^{2}\right)
\ ,
\label{metric}
\end{equation}
where $\nu =\nu (r)$ and $\lambda =\lambda (r)$ are functions of the areal
radius $r$ only, ranging from $r=0$ (the star center) to some $r=R$ (the
star surface). The metric~(\ref{metric}) must satisfy the Einstein equations~(\ref{corr2}),
which in terms of the two sources in~\eqref{emt} explicitly read, 

\begin{eqnarray}
\label{ec1}
k^2
\left(
T_0^{\ 0}+\theta_0^{\ 0}
\right)
&\!\!=\!\!&
\frac 1{r^2}
-
e^{-\lambda }\left( \frac1{r^2}-\frac{\lambda'}r\right)\ ,
\\
\label{ec2}
k^2
\left(T_1^{\ 1}+\theta_1^{\ 1}\right)
&\!\!=\!\!&
\frac 1{r^2}
-
e^{-\lambda }\left( \frac 1{r^2}+\frac{\nu'}r\right)\ ,
\\
\label{ec3}
k^2
\left(T_2^{\ 2}+\theta_2^{\ 2}\right)
&\!\!=\!\!&
\frac {e^{-\lambda }}{4}
\left(-2\nu''-\nu'^2+\lambda'\nu'
-2\,\frac{\nu'-\lambda'}r\right)
\ ,
\nonumber \\
\end{eqnarray}
where $f'\equiv \partial_r f$ and $\tilde{T}_3^{{\ 3}}=\tilde{T}_2^{\ 2}$ due to the spherical symmetry.
The conservation equation~(\ref{dT0}) is a linear combination of Eqs.~(\ref{ec1})-(\ref{ec3}), and yields
\begin{equation}
\label{con1}
\left(\tilde{T}_1^{\ 1}\right)'
-
\frac{\nu'}{2}\left(\tilde{T}_0^{\ 0}-\tilde{T}_1^{\ 1}\right)
-
\frac{2}{r}\left(\tilde{T}_2^{\ 2}-\tilde{T}_1^{\ 1}\right)
=
0
\ ,
\end{equation}
which in terms of the two sources in Eq.~\eqref{emt} read,
\begin{eqnarray}
\label{con11}
&&\left({T}_1^{\ 1}\right)'
-
\frac{\nu'}{2}\left({T}_0^{\ 0}-{T}_1^{\ 1}\right)
-
\frac{2}{r}\left({T}_2^{\ 2}-{T}_1^{\ 1}\right)
\nonumber\\
&&+\left({\theta}_1^{\ 1}\right)'
-
\frac{\nu'}{2}\left({\theta}_0^{\ 0}-{\theta}_1^{\ 1}\right)
-
\frac{2}{r}\left({\theta}_2^{\ 2}-{\theta}_1^{\ 1}\right)
=
0 \ .
\nonumber\\
\end{eqnarray}
\par
By simple inspection, we can identify in Eqs.~\eqref{ec1}-\eqref{ec3} an effective density  
\begin{equation}
\tilde{\rho}
=
T_0^{\ 0}
+\theta_0^{\ 0}
\ ,
\label{efecden}
\end{equation}
an effective radial pressure
\begin{equation}
\tilde{p}_{r}
=
-T_1^{\ 1}-\theta_1^{\ 1}
\ ,
\label{efecprera}
\end{equation}
and an effective tangential pressure
\begin{equation}
\tilde{p}_{t}
=
-T_2^{\ 2}-\theta_2^{\ 2}
\ .
\label{efecpretan}
\end{equation}
These definitions clearly illustrate  an
anisotropy
\begin{equation}
\label{anisotropy}
\Pi
\equiv
\tilde{p}_{t}-\tilde{p}_{r}
\ .
\end{equation}
The system of Eqs.~(\ref{ec1})-(\ref{ec3}) may therefore be formally
treated as an anisotropic fluid~\cite{Luis,tiberiu}.
Now let us implement the MGDe to solve the system Eqs.~(\ref{ec1})-(\ref{con1}). We will see that under this approach, the system will be transformed in such a way that the equation of motions associated with the source $\theta_{\mu\nu}$ will satisfy an effective ``quasi-Einstein system" [see further Eqs. (\ref{ec1d})-(\ref{ec3d})].
\par
\par
We can then proceed by considering a solution to the Eqs.~\eqref{corr2} for the source $T_{\mu\nu}$ [that is Eqs.~(\ref{ec1})-(\ref{con1}) with $\theta_{\mu\nu}=0$], which we can write as
\begin{equation}
ds^{2}
=
e^{\xi (r)}\,dt^{2}
-e^{\mu (r)}\,dr^{2}
-
r^{2}\left( d\theta^{2}+\sin ^{2}\theta \,d\phi ^{2}\right)
\ ,
\label{pfmetric}
\end{equation}
where 
\begin{equation}
\label{standardGR}
e^{-\mu(r)}
\equiv
1-\frac{k^2}{r}\int_0^r x^2\,T_0^{\, 0}(x)\, dx
=
1-\frac{2\,m(r)}{r}
\end{equation}
is the standard General Relativity expression containing the Misner-Sharp mass function $m(r)$.
The effects of the source $\theta_{\mu\nu}$ on the source $T_{\mu\nu}$ can then
be encoded in the geometric deformation undergone by the $T_{\mu\nu}$ geometry $\{\xi,\,\mu\}$ in Eq.~(\ref{pfmetric}), namely
\begin{eqnarray}
\label{gd1}
\xi &\rightarrow &\nu\,=\,\xi+\alpha\,g\ ,
\\
\label{gd2}
e^{-\mu} &\rightarrow &e^{-\lambda}=e^{-\mu}+\alpha\,f , 
\end{eqnarray}
where $f$ and $g$ are respectively the geometric deformations undergone by the radial and temporal metric component and the constant $\alpha$ a free parameter. Now let us plug the  decomposition in Eqs.~(\ref{gd1}) and (\ref{gd2}) in the Einstein equations~(\ref{ec1})-(\ref{ec3}). The system is thus separated in two sets: i) one having the standard Einstein field equations for an energy-momentum tensor $T_{\mu\nu}$, whose metric is given by Eq.~(\ref{pfmetric})
\begin{eqnarray}
\label{ec1pf}
&&
k^2\,T_0^{\, 0}
=\frac 1{r^2}
-
e^{-\mu }\left( \frac1{r^2}-\frac{\mu'}r\right)\ ,
\\
&&
\label{ec2pf}
k^2
\,T_1^{\, 1}
=
\frac 1{r^2}
-
e^{-\mu}\left( \frac 1{r^2}+\frac{\xi'}r\right)\ ,
\\
&&
\label{ec3pf}
k^2
\strut\displaystyle
\,T_2^{\, 2}
=\frac {e^{-\mu }}{4}
\left(-2\xi''-\xi'^2+\mu'\xi'
-2\,\frac{\xi'-\mu'}r\right)
\ ,
\end{eqnarray}
and ii) one with the equation of motion for the source $\theta_{\mu\nu}$, which reads
\begin{eqnarray}
\label{ec1d}
&&
k^2\,\theta_0^{\ 0}
=-\frac{\alpha\,f}{r^2}
-\frac{\alpha\,f'}{r}\ ,
\\
&&
\label{ec2d}
k^2\,\theta_1^{\ 1}+\alpha\,Z_1
= -\alpha\,f\left(\frac{1}{r^2}+\frac{\nu'}{r}\right)\ ,
\\
&&
\label{ec3d}
k^2\,\theta_2^{\ 2}+\alpha\,Z_2
=-\frac{\alpha\,f}{4}\left(2\,\nu''+\nu'^2+2\frac{\nu'}{r}\right)
\nonumber \\ &&
\,\,\,\,\,\,\,\,\,\,\,\,\,\,\,\,\,\,\,\,\,\,\,\,\,\,\,\,\,\,\,\,\,\,\,\,\,\,\,\,\,\,\,\,\,\,\,-\frac{\alpha\,f'}{4}\left(\nu'+\frac{2}{r}\right)
\ ,
\end{eqnarray}
where $Z_1$ and $Z_2$ are defined as
\begin{eqnarray}
\label{Z1}
Z_1&=&\frac{e^{-\mu}\,g'}{r}\ ,
\\
\label{Z2}
4\,Z_2&=&e^{-\mu}\left(2g''+\alpha\,g'^2+\frac{2\,g'}{r}+2\xi'\,g'-\mu'g'\right)\ ,
\end{eqnarray}
while the conservation equation in~\eqref{con11} becomes
\begin{eqnarray}
\label{con111}
&&\left[\left({T}_1^{\ 1}\right)'
-
\frac{\xi'}{2}\left({T}_0^{\ 0}-{T}_1^{\ 1}\right)
-
\frac{2}{r}\left({T}_2^{\ 2}-{T}_1^{\ 1}\right)\right]
\nonumber \\
&&-\frac{\alpha\,g'}{2}\left({T}_0^{\ 0}-{T}_1^{\ 1}\right)
\nonumber\\
&&+\left({\theta}_1^{\ 1}\right)'
-
\frac{\nu'}{2}\left({\theta}_0^{\ 0}-{\theta}_1^{\ 1}\right)
-
\frac{2}{r}\left({\theta}_2^{\ 2}-{\theta}_1^{\ 1}\right)
=
0\ ,
\nonumber \\
\end{eqnarray}
where the bracket in Eq.~\eqref{con111} represents the divergence of $T_{\mu\nu}$ calculated with the metric in Eq.~(\ref{pfmetric}). We can see that in the particular case of the MGD, namely $g'(r)=0$, the system~\eqref{ec1d}-\eqref{ec3d} is reduced to the ``quasi-Einstein" system found in Ref.~\cite{MGD-decoupling}. Next we shall  discuss a critical point regarding the conservation equation~\eqref{con111} in order to have a successful decoupling. 
\par
Since the Einstein tensor ${G}_{\mu\nu}$ associated with the $T_{\mu\nu}$ geometry $\{\xi,\,\mu\}$ in Eq.~(\ref{pfmetric}) must satisfy its respective Bianchi identity, the energy momentum tensor $T_{\mu \nu }$ is conserved under this geometry, explicitly shown in Eqs.~\eqref{ec1pf}-\eqref{ec3pf}, so that
\begin{equation}
\label{pfcon}
\nabla^{(\xi,\mu)}_\sigma\,T^{\sigma}_{\, \nu}=0\ ,
\end{equation}
where $\nabla^{(\xi,\mu)}$ means that the divergence in Eq.~\eqref{pfcon} is calculated with the metric in Eq.~(\ref{pfmetric}). Notice that,
\begin{equation}
\label{divs}
\nabla_\sigma\,T^{\sigma}_{\, \nu}=\nabla^{(\xi,\mu)}_\sigma\,T^{\sigma}_{\, \nu}-\frac{\alpha\,g'}{2}\left({T}_0^{\ 0}-{T}_1^{\ 1}\right)\delta^1_\nu\ ,
\end{equation}
where the divergence in the left-hand side in Eq.~\eqref{divs} is calculated with the metric in Eq.~\eqref{metric}. As expected, the expression~\eqref{pfcon}, which explicitly read
\begin{equation}
\label{pfcon2}
\left({T}_1^{\ 1}\right)'
-
\frac{\xi'}{2}\left({T}_0^{\ 0}-{T}_1^{\ 1}\right)
-
\frac{2}{r}\left({T}_2^{\ 2}-{T}_1^{\ 1}\right)=0\ ,
\end{equation}
is a linear combination of the Einstein field equations~\eqref{ec1pf}-\eqref{ec3pf}. Thus we can assure that the source $T_{\mu\nu}$ has been successfully decoupled from the system~\eqref{ec1}-\eqref{con1}.
\par
Finally, taking into account the condition~\eqref{pfcon} and the energetic balance in Eq.~\eqref{con111}, we have
\begin{equation}
\label{exch1}
\nabla_\sigma\,T^{\sigma}_{\, \nu}=-\frac{\alpha\,g'}{2}\left({T}_0^{\ 0}-{T}_1^{\ 1}\right)\delta^1_\nu
\end{equation}
and
\begin{equation}
\label{exch2}
\nabla_\sigma\theta^{\sigma}_{\nu}=\frac{\alpha\,g'}{2}\left({T}_0^{\ 0}-{T}_1^{\ 1}\right)\delta^1_\nu\ ,
\end{equation}
where the divergence in Eqs.~\eqref{exch1} and~\eqref{exch2} is calculated with the metric in Eq.~(\ref{metric}). The expression in Eq.~\eqref{exch2} explicitly read
\begin{eqnarray}
\label{con22}
\left({\theta}_1^{\ 1}\right)'
-
\frac{\nu'}{2}\left({\theta}_0^{\ 0}-{\theta}_1^{\ 1}\right)
-
\frac{2}{r}\left({\theta}_2^{\ 2}-{\theta}_1^{\ 1}\right)
=
\frac{\alpha\,g'}{2}\left({T}_0^{\ 0}-{T}_1^{\ 1}\right)
\ ,
\nonumber \\
\end{eqnarray}
which is a linear combination of the equation of motions for the source $\theta_{\mu\nu}$, displayed in the expressions~\eqref{ec1d}-\eqref{ec3d}. We therefore conclude that both sources $T_{\mu\nu}$ and $\theta_{\mu\nu}$ can be successfully decoupled as long as there is exchange of energy between them, as we can see in the expressions~\eqref{exch1} and~\eqref{exch2}. We see the particular case $g=0$, namely the MGD-decoupling in Ref.~\cite{MGD-decoupling}, allows a decoupling without exchange of energy between the sources. Also is worth noting that a successful decoupling without exchange of energy is possible under MGDe when i) $T_{\mu\nu}$ is a barotropic fluid whose equation of state is ${T}_0^{\ 0} = {T}_1^{\ 1}$; and ii) for those regions where $T_{\mu\nu}=0$ \footnote{This explains the successful extension of the MGD, valid only in the vacuum, found in Ref.~\cite{MGDextended1}}. The latter can be the case of a region $r>R$ filled by a source $\theta_{\mu\nu}$ surrounding a self-gravitating system of radius $R$ and source $T_{\mu\nu}$. 
\par
Finally, we remark that the equation of motions for the source $\theta_{\mu\nu}$ in Eqs.~\eqref{ec1d}-\eqref{ec3d}
may be formally identified as Einstein equations for an anisotropic system with a {\it conserved} energy-momentum tensor ${\theta}^{*}_{\mu\nu}$ defined as
\begin{eqnarray}
\label{shift2m}
k^2\,\theta_\mu^{*\,\,\nu}&=&k^2\,\theta_\mu^{\,\nu}+\frac{1}{r^2}\delta_\mu^{\,\,\,0}\,\delta_0^{\,\,\,\nu}+\left(\alpha\,Z_1+\frac{1}{r^2}\right)\delta_\mu^{\,\,\,1}\,\delta_1^{\,\,\,\nu}
\nonumber \\
&&+\alpha\,Z_2\left(\delta_\mu^{\,\,\,2}\,\delta_2^{\,\,\,\nu}+\delta_\mu^{\,\,\,3}\,\delta_3^{\,\,\,\nu}\right)\ ,
\end{eqnarray}
with the conservation equation 
\begin{eqnarray}
\label{con222}
\left({\theta^*}_1^{\ 1}\right)'
-
\frac{\nu'}{2}\left({\theta^*}_0^{\ 0}-{\theta^*}_1^{\ 1}\right)
-
\frac{2}{r}\left({\theta^*}_2^{\ 2}-{\theta^*}_1^{\ 1}\right)
= 0
\end{eqnarray}
and metric
\begin{equation}
ds^{2}
=
e^{\nu(r)}\,dt^{2}
-
\frac{dr^{2}}{f(r)}
-
r^{2}\left( d\theta^{2}+\sin ^{2}\theta \,d\phi ^{2}\right)
\ .
\label{thetametric}
\end{equation}
\par
We conclude we have successfully decoupled Einstein equations in Eqs.~(\ref{ec1})-(\ref{ec3}) in two systems, namely: i) Einstein field equations for a source $T_{\mu\nu}$ in Eqs~(\ref{ec1pf})-(\ref{ec3pf}) to determine $\{T_{\mu\nu},\,\xi,\,\mu\}$, and ii) field equations for the source $\theta_{\mu\nu}$ in Eqs~(\ref{ec1d})-(\ref{ec3d}) to determine $\{\theta_{\mu\nu},\,g,\,f\}$; or equivalently, Einstein field equations for the conserved source $\theta^{*}_{\mu\nu}$ in Eq.~\eqref{shift2m} to determine $\{\theta^{*}_{\mu\nu},\,g,\,f\}$. 
\par
At this stage the reader may be tempted to ensure that it is easier to work with the well-known original system in Eqs.~(\ref{ec1})-(\ref{ec3}), instead of using the unfamiliar gravitational system displayed in Eqs.~(\ref{ec1d})-(\ref{ec3d}), since the former has eight unknowns $\{T_{\mu\nu},\,\theta_{\mu\nu},\,\nu,\,\lambda\}$, thus leaving $8-3=5$ degrees of freedom to be used in the search for a solution, rather than the five unknowns $\{\theta_{\mu\nu},\,\xi,\,g\}$ and only $5-3=2$ degrees of freedom for a new and unexplored set of equations. However, we must keep in mind that the goal of MGDe is to separate the gravitational sources in such a way that we can focus exclusively on each of them, in a systematic way, either i) to study a complex gravitational source by its deconstruction in its simplest components, namely, $\tilde{T}_{\mu\nu}\rightarrow\{T_{\mu\nu},\,\theta_{\mu\nu}\}$ , or ii) the extension of a simple gravitational source to more complex forms of the energy-momentum tensor, namely, $T_{\mu\nu}\rightarrow\,\tilde{T}_{\mu\nu}={T}_{\mu\nu}+\theta_{\mu\nu}$, as those developed in Refs.~\cite{mgdaniso,Luciano,MGDDBH,sharif1,Inverse,sharif2,ernestopedro,tello,camilo,sharif3}. Indeed, this last point of view is particularly useful when we face alternative theories, which usually lead to equations difficult to deal with. In this respect, let us consider the following modified Einstein-Hilbert action
\begin{equation}
\label{ngt}
S_{\rm G}=S_{\rm EH}+S_{\rm X}=\int\left[\frac{R}{2\,k^2}+{\cal L}_{\rm M}+{\cal L}_{\rm X}\right]\sqrt{-g}\,d^4\,x\ ,
\end{equation}
where $R$ is the Ricci scalar, ${\cal L}_{\rm M}$ contains any matter fields appearing in the theory and ${\cal L}_{\rm X}$ the Lagrangian density of a new gravitational sector not described by general relativity, let us say a ``${\rm X}$-gravitational sector". It could be, among others, a Lovelock's theory of gravity. This new sector always can be seen as corrections to general relativity and be consolidated as part of an effective energy-momentum tensor $\theta_{\mu\nu}$, definite as
\begin{equation}
\label{ngt2}
\theta_{\mu\nu}=\frac{2}{\sqrt{-g}}\frac{\delta\,(\sqrt{-g}\,{\cal L}_{\rm X})}{\delta\,g^{\mu\nu}}=2\,\frac{\delta\,{\cal L}_{\rm X}}{\delta\,g^{\mu\nu}}-\,g_{\mu\nu}{\cal L}_{\rm X}\ ,
\end{equation}
just like that displayed in Eq.~\eqref{emt}. The advantage of MGDe-decoupling now becomes quite clear: we can extend all the known solutions associated with the action $S_{\rm EH}$, namely, solutions $\{T_{\mu\nu},\,\xi,\,\mu\}$ of the system displayed in Eqs~(\ref{ec1pf})-(\ref{ec3pf}), into the domain of modified gravity represented by $S_{\rm G}$ [and its equation of motions in Eqs.~(\ref{ec1})-(\ref{ec3})], by solving the unconventional gravitational system displayed in Eqs.~(\ref{ec1d})-(\ref{ec3d}) to determine $\{\theta_{\mu\nu},\,g,\,f\}$. Hence we can generate the ``${\rm X}$-version" of any $\{T_{\mu\nu},\,\xi,\,\mu\}$-solution, namely,
\begin{equation}
\label{modified}
\{T_{\mu\nu},\,\xi,\,\mu\}\Rightarrow\,\{\tilde{T}_{\mu\nu},\,\nu,\,\lambda\}\ .
\end{equation}
The above represents a straightforward way to study the consequences of extended gravity on general relativity. 
\section{Decoupling Einstein-Maxwell}
\label{s3}
\par
With the aim of not only testing the consistency of MGDe, but also to show its advantages and the way it works, we shall consider a well-known case, namely the Einstein-Maxwell system.
\par
For a charged self-gravitating distribution, the generic source $\theta_{\mu\nu}$ in Eq.~\eqref{emt} takes the form of the Maxwell energy-momentum tensor, given by
\begin{equation}
\label{max}
\theta_{\mu\nu}=\frac{1}{4\pi}\left[F_{\mu\alpha}F^{\alpha}_{\,\,\,\,\nu}+\frac{1}{4}g_{\mu\nu}F_{\alpha\beta}F^{\alpha\beta}\right]\ .
\end{equation}
We just recall that the electromagnetic field $F_{\mu\nu}$ in Eq.~(\ref{max}) satisfies Maxwell's equations
\begin{eqnarray}
\label{me}
\nabla_\nu\left[(-g)^{1/2}F^{\mu\nu}\right]&=&4\pi\,(-g)^{1/2}j^{\mu}\ ,
\\
\partial_{[\sigma}F_{\mu\nu]}&=&0\ ,
\end{eqnarray}
where $j^{\mu}$ is the four-current, which in the static case becomes $j^{\mu}=(j^0,0,0,0)$. Because of the spherical symmetry, only the radial electric field $F^{01}=-F^{10}$ is non-vanishing, and given by
\begin{equation}
\label{F01}
F^{01}=\frac{e^{-(\nu+\lambda)/2}\,q(r)}{r^2}\ ,
\end{equation}
where $q(r)$ is the electric charge~\cite{Bekenstein71} of a spherical system of radius $r$, defined as
\begin{equation}
\label{Q}
q(r)=\int_0^r\,4\pi\,x^2e^{\frac{\nu+\lambda}{2}}\,j^0\,dx\ .
\end{equation}
Using Eqs.~\eqref{max} and~\eqref{F01}, the system~(\ref{ec1d})-({\ref{ec3d}}) becomes
\begin{eqnarray}
\label{ec1m}
&&
{\cal E}^2
=
-\strut\displaystyle\frac{\alpha\,f}{r^2}
-\frac{\alpha\,f'}{r}\ ,
\\
&&
\label{ec2m}
{\cal E}^2+\alpha\,Z_1
=-\alpha\,f\left(\frac{1}{r^2}+\frac{\nu'}{r}\right)\ ,
\\
&&
\label{ec3m}
-{\cal E}^2+\alpha\,Z_2
=-\frac{\alpha\,f}{4}\left(2\,\nu''+\nu'^2+2\frac{\nu'}{r}\right)
\nonumber \\
&&\,\,\,\,\,\,\,\,\,\,\,\,\,\,\,\,\,\,\,\,\,\,\,\,\,\,\,\,\,\,\,\,\,\,\,\,\,\,\,-\frac{\alpha\,f'}{4}\left(\nu'+\frac{2}{r}\right)
\ ,
\end{eqnarray}
with the expression in Eq.~(\ref{con22}) yields
\begin{equation}
\label{conmax}
({\cal E}^2)'+\frac{4{\cal E}^2}{r}=\frac{\alpha\,g'}{2\,r}\left(\mu\,\xi'-\mu'\right)\ ,
\end{equation}
which is a linear combination of Eqs.~\eqref{ec1m}-\eqref{ec3m}, and where ${\cal E}=q/r^2$ is the electric field intensity. We shall stop here for a moment to emphasize what we have done so far:
\begin{itemize}

\item First of all, we have started from an electrically charged distribution represented by the system displayed in Eqs.~\eqref{ec1}-\eqref{con11}, where the source $\theta_{\mu\nu}$ has the form in Eq.~\eqref{max}. 

\item Next, by using the MGDe-decoupling, we have ``extracted" the generic source $T_{\mu\nu}$ from Eqs.~\eqref{ec1}-\eqref{con11} in such a way that it still satisfies Einstein's field equations, but now with a different geometry $\{\nu\rightarrow\xi,\,e^{-\lambda}\rightarrow\,e^{-\mu}\}$ displayed in Eqs.~(\ref{ec1pf})-(\ref{ec3pf}), thus leaving everything concerning the charged source in the unconventional Einstein-Maxwell system~\eqref{ec1m}-\eqref{conmax}.

\end{itemize}
The above clearly shows the simplification introduced by MGDe-decoupling: if we already have a solution $\{T_{\mu\nu},\,\xi,\,\mu\}$ for the system~(\ref{ec1pf})-(\ref{ec3pf}), then we can use this information to feed the second  system~\eqref{ec1m}-\eqref{conmax}. Hence we end with three unknown functions $\{{\cal E},f,g\}$ which can be found by three of the equations in Eqs.~(\ref{ec1m})-(\ref{conmax}). Therefore we can generate, in a straightforward way, the ``charged version" of the solution $\{T_{\mu\nu},\,\xi,\,\mu\}$. Of course we can repeat this procedure for any other theory represented by the action $S_{\rm X}$ in Eq.~\eqref{ngt}, whose effective energy-momentum tensor is displayed in Eq.~\eqref{ngt2}. 
\par
Now let us consider the simplest situation, namely, the vacuum $T_{\mu\nu}=0$ for the region $r>R$, where $r=R$ is the surface of the self-gravitating system. Therefore, in terms of the MGDe nomenclature, we shall build the ``Maxwell version" of the vacuum $T_{\mu\nu}=0$. In this region the metric functions $\xi$ and $\mu$ in Eq.~(\ref{pfmetric}) are given by the Schwarzschild solution, which is ``deformed" according to Eqs.~(\ref{gd1}) and (\ref{gd2}), yielding thus to a new solution given by.
\begin{eqnarray}
\label{Schwdef}
ds^{2}
=&&\left(1-\frac{2\,M}{r}\right)\,e^{\alpha\,g(r)}\,dt^{2}
-\left(1-\frac{2\,M}{r}+\alpha\,f(r)\right)^{-1}dr^2
\nonumber\\
&&-r^{2}\left(d\theta ^{2}+\sin {}^{2}\theta d\phi ^{2}\right)
\ .
\end{eqnarray}
Since for the Schwarzschild solution $e^\xi=\mu$, the right-hand side in Eq.~(\ref{conmax}) vanishes, thus the electric field intensity is found as  
\begin{equation}
\label{E}
{\cal E}=\frac{Q}{r^2}\ ,
\end{equation}
where $Q$ is a constant which eventually is identified as the total electric charge. By using Eq.~(\ref{E}) in Eq.~(\ref{ec1m}) we obtain $f(r)$ as
\begin{equation}
\label{fm}
\alpha\,f(r)=\frac{c_1}{r}+\frac{Q^2}{r^2}\ ,
\end{equation}
where $c_1$ is a constant. Combining Eqs.~(\ref{ec1m}) and (\ref{ec2m}) we obtain
\begin{equation}
\alpha\,g'(r)=\frac{\alpha\,f'-\alpha\,f\,\xi'}{\mu+\alpha\,f}
\end{equation}
which can easily be integrated, yielding 
\begin{equation}
\label{g}
e^{\alpha\,g}=c_2\left(1-\frac{2\,M}{r}\right)^{-1}\left(1-\frac{2\,M}{r}+\frac{c_1}{r}+\frac{Q^2}{r^2}\right)\ ,
\end{equation}
where $c_2$ is an integration constant which can be taken as $1$ without loss of generality [equivalent to the time transformation $dT=c_2\,dt$ in Eq.~\eqref{Schwdef}]. Using the expressions in Eq.~(\ref{fm}) and Eq.~(\ref{g}) in the metric shown in Eq.~(\ref{Schwdef}), we obtain
\begin{eqnarray}
ds^{2}
=&&\left(1-\frac{2\,{ M}}{r}+\frac{Q}{r^2}\right)\,dt^{2}
-\left(1-\frac{2\,{ M}}{r}+\frac{Q}{r^2}\right)^{-1}dr^2
\nonumber\\
&&-r^{2}\left(d\theta ^{2}+\sin {}^{2}\theta d\phi ^{2}\right)
\ ,
\label{RN}
\end{eqnarray}
where we have taken $c_1=0$ (we can also define ${\cal M}\equiv\,M-\frac{c_1}{2}$). This is, as expected, the well-known Reissner-Nordstrom solution.
%
%
%
%
%
%
%
\section{Conclusions}
\label{con}
\setcounter{equation}{0}
By making use of the MGD-decoupling approach, we have presented in detail the most 
general way to decoupling two spherically symmetric and static gravitational sources $\{T_{\mu\nu},\,\theta_{\mu\nu}\}$ in general relativity, namely, i) when both metric functions $\{g_{tt},\,g_{rr}\}$ are deformed; and ii) valid for all regions of the space-time, regardless of whether there is matter or not. The above was achieved through a careful and detailed analysis of Bianchi's identities, which shows that a successful decoupling of the Einstein field equations~\eqref{ec1}-\eqref{ec3} is only possible through a specific exchange of energy between both sources, as displayed in Eqs.~\eqref{exch1} and~\eqref{exch2}. The decoupling of these gravitational sources yields two systems, namely, Einstein field equations for the source $T_{\mu\nu}$, displayed in Eqs.~\eqref{ec1pf}-\eqref{ec3pf}, and the ``quasi-Einstein'' system for the source $\theta_{\mu\nu}$, shown in Eqs.~\eqref{ec1d}-\eqref{ec3d}. Both systems are complemented with their own conservation equation, displayed in Eqs.~\eqref{pfcon2} and~\eqref{con22} respectively. Also we found that a successful decoupling without exchange of energy is possible as long as it occurs in the ``vacuum'' $T_{\mu\nu}=0$, or when ${T}_0^{\ 0} = {T}_1^{\ 1}$. In these two cases the interaction between both sources is purely gravitational.~\footnote{By ``vacuum'' we mean those regions where one source is zero, or more properly, those regions where the sources do not coexist}
\par
We also show the potential of the MGDe-decoupling to investigate the consequences of modified gravity on general relativity. In this respect, the new gravitational sector, generically represented by $S_{\rm X}$ in the action~\eqref{ngt}, and whose effective energy-momentum tensor is displayed in Eq.~\eqref{ngt2}, may represent a huge range of alternative theories. Hence, following the MGDe-decoupling, we can separate the pure Einstein sector, namely $S_{\rm EH}$ in Eq.~\eqref{ngt}, from the modified Einstein-Hilbert action $S_{\rm G}$ [indeed, this is what we have done when we ``extract'' the system~\eqref{ec1pf}-\eqref{ec3pf} from Eqs.~\eqref{ec1}-\eqref{ec3}]. This turns out to be tremendously useful if we want to extend a solution from the pure Einstein sector into the domain of modified gravity, represented by $S_{\rm G}$, or equivalently, to elucidate the effects of a new gravitational sector, represented by $S_{\rm X}$, on general relativity.
\par
In order to examine the consistency of the MGDe-decoupling, we have applied it to the well-known Einstein-Maxwell system, represented by two sources $\{T_{\mu\nu},\,F_{\mu\nu}\}$. Following the MGDe-decoupling, we separated the Einstein-Maxwell field equations into i) pure Einstein for the source $T_{\mu\nu}$ and ii) the “quasi-Einstein” system~\eqref{ec1d}-\eqref{ec3d}, which leads to an unconventional Einstein-Maxwell system displayed in Eqs.~\eqref{ec1m}-\eqref{conmax}. Then we considered the simplest case, namely, the vacuum $T_{\mu\nu}=0$, where the MGDe-decoupling shows the effects of the Maxwell theory on the spherically symmetric gravitational vacuum in terms of the deformed Schwarzschild metric~\eqref{Schwdef}, whose explicit form is nothing but the well-known Reissner-Nordstrom solution. 
\par
We conclude by highlighting some aspects regarding the “quasi-Einstein” system~\eqref{ec1d}-\eqref{ec3d}. First of all, since the vacuum $T_{\mu\nu}=0$ leads to a pure gravitational interaction between the sources, and since the interstellar regions have low energy densities $\rho\sim\,p\sim\,0$, we conclude that the quasi-Einstein system represents a good candidate to investigate the conjectured dark matter. In fact, whether it exists or is a consequence of some modified gravity, dark matter can always be represented by the energy-momentum tensor $\theta_{\mu\nu}$ in Eqs.~\eqref{ec1d}-\eqref{ec3d}, which is the source of the deformation $\{g,\,f\}$ undergone by the Schwarzschild solution, explicitly display in Eq.~\eqref{Schwdef}. Therefore, by introducing some physically reasonable information to the system~\eqref{ec1d}-\eqref{ec3d}, we could reproduce some phenomena associated with dark matter, and more importantly, predict new phenomena associated with it. A second aspect to highlight is the potential of the system~\eqref{ec1d}-\eqref{ec3d} to investigate hairy black holes~\cite{thomas1,kanti1,kanti2,kanti3}. We can see this clearly when a source $\theta_{\mu\nu}$ fills the spherically symmetric gravitational vacuum $T_{\mu\nu}=0$. Hence its generic solution will be that displayed in Eq.~\eqref{Schwdef}. Finally, we conclude by rising some natural questions regarding the MGDe-decoupling which deserve to be investigated, such as: its validity for time-dependent configurations~\cite{static}, its possible extension beyond the spherical symmetry, and its generalization for extra-dimensional space-times.  


%
%
%
%
\section{Acknowledgements}
\par
This work has been supported by the Albert Einstein Centre for Gravitation and Astrophysics financed
by the Czech Science Agency Grant No.14-37086G.
%

%
%


\begin{thebibliography}{99}
\bibitem{static} In the non-static case, the component of the Einstein tensor $G_2^{\,\,2}$ in~\eqref{ec2} has the additional terms $\sim\ddot{\lambda}+\dot{\lambda}(\dot{\lambda}-\dot{\nu})$ which makes impossible a decoupling under the MGDe-transformation~\eqref{gd1} and~\eqref{gd2}. Therefore a non-trivial extension is necessary.
\end{thebibliography}
\end{document}